\documentstyle[pre,aps,amsmath,epsfig]{revtex}

\def\vec#1{\mbox{\boldmath{$#1$}}}

\begin{document}

\twocolumn[%
\hsize\textwidth\columnwidth\hsize\csname@twocolumnfalse\endcsname
\title{Granular cooling of hard needles}
\author{Martin Huthmann, Timo Aspelmeier, and Annette Zippelius}
\address{Institut f\"ur Theoretische Physik, 
         Universit\"at G\"ottingen, 
         D-37073 G\"ottingen, Germany}
\date{\today}
\maketitle
\begin{abstract}
  We have developed a kinetic theory of hard needles undergoing binary
  collisions with loss of energy due to normal and tangential
  restitution.  In addition, we have simulated many particle systems
  of granular hard needles.  The theory, based on the assumption of a
  homogeneous cooling state, predicts that granular cooling of the
  needles proceeds in two stages: An exponential decay of the initial
  configuration to a state where translational and rotational energies
  take on a time independent ratio (not necessarily unity), followed
  by an algebraic decay of the total kinetic energy $\sim t^{-2}$.
  The simulations support the theory very well for low and moderate
  densities.  For higher densities, we have observed the onset of the
  formation of clusters and shear bands.
\end{abstract}
\vspace*{-2.5pt}
]

\section{Introduction}
Gas kinetics of inelastically colliding particles has received a lot
of interest in recent years, mainly in the context of granular matter
\cite{NATOASI98}.  Boltzmann's equation has been generalized to
inelastic collisions which are characterized by normal restitution and
possibly tangential friction or restitution \cite{boltzies}.
Extensive simulations have been performed, in particular event driven
(ED) algorithms are very effective for the kinetic gas regime.  A
variety of interesting phenomena have been observed, for example
instabilities of the homogeneous state towards shearing or clustering
\cite{goldhirsch93,mcnamara96,noije97}.  Most studies so far have
concentrated on spherically symmetric objects, whereas real grains are
in general nonspherical and often randomly shaped. The question arises
whether the observed phenomena are generic for granular matter or
model specific.

In this paper we discuss the cooling properties of hard needles in
terms of a time evolution operator, which accounts for the exchange of
translational and rotational energy as well as for normal and
tangential restitution. In addition we have performed ED simulations
of large systems with up to 20000 needles.

For low and moderate densities the system does not show any clustering
instabilities but remains homogeneous on the longest time scales, when
the energy has decayed to $10^{-11}$ of its initial value.  This
allows us to formulate an approximate kinetic theory, based on the
assumption of a homogeneous state. Cooling is found to proceed in two
stages: 1) A fast exponential decay to a state which is characterized
by a time independent ratio $c$ of translational to rotational energy.
2) A slow algebraic decay like $t^{-2}$ of the total kinetic energy,
while the ratio $c$ exhibits small fluctuations around its time
independent mean value. The latter is determined by the coefficient of
restitution and by the distribution of mass along the needles,
including equipartition for one particular mass distribution.
Simulations and approximate analytical theory are found to agree
within a few percent.

For high densities we observe large scale structures in the velocity
field, similar to those seen in dense systems of smooth spheres
\cite{goldhirsch93,mcnamara96,noije97,noije97.2}. These structures
give rise to long range correlations in the velocity field and the
build up of large density fluctuations. No alignment of the needles
with the hydrodynamic flow field is observed.

\section{Dynamics of collisions}
\begin{figure}[htbp]
  \begin{center}
    \leavevmode
    \epsfig{file=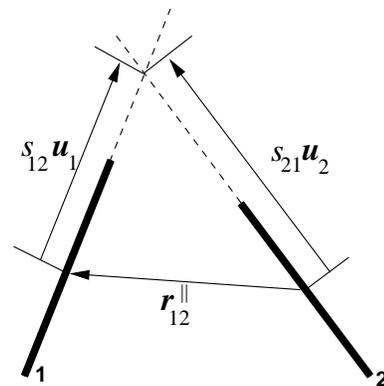, width=5cm}
    \caption{Configuration of two needles projected in the plane
      $E_{12}$ spanned by the unitvectors $\vec{u}_1$ and $\vec{u}_2$}
\label{bild1}
  \end{center}
\end{figure}

Consider two rods of equal length $L$, mass $m$ and moment of inertia
$I$. The center of mass coordinates are denoted by $\vec{r}_1$ and
$\vec{r}_2$. The orientations are specified by unit vectors
$\vec{u}_1$ and $\vec{u}_2$, which span a plane $E_{12}$ with normal
\begin{equation}
\vec{u}_\perp =
  \frac{\vec{u}_1 \times \vec{u}_2} {|\vec{u}_1 \times
  \vec{u}_2|}.
\end{equation}
We decompose $\vec{r}_{12}=\vec{r}_1-\vec{r}_2$ into a component
perpendicular $\vec{r}_{12}^\perp =
(\vec{r}_{12}\cdot\vec{u}_\perp)\vec{u}_\perp$ and parallel
$\vec{r}_{12}^{\parallel} =: (s_{12}\vec{u}_1-s_{21}\vec{u}_2) $ to
$E_{12}$ (see Fig. \ref{bild1}).  The rods are in contact
\cite{frenkel83,cichocki86} if $\vec{r}_{12}^\perp=0$ and
simultaneously $|s_{12}|<L/2$ and $|s_{21}|<L/2$.

We want to determine the postcollisional center-of-mass velocities
($\vec{v}'_1$, $\vec{v}'_2$) and angular velocities
($\vec{\omega}'_1$, $\vec{\omega}'_2$) in terms of the precollisional
velocities ($\vec{v}_1$, $\vec{v}_2$, $\vec{\omega}_1$, and
$\vec{\omega}_2$) or momenta
$(\vec{p}_1=m\vec{v}_1,\vec{p}_2=m\vec{v}_2)$.  Conservation of total
linear momentum implies
\begin{align} \nonumber
   \vec{p}'_1 &= \vec{p}_1 +\Delta \vec{p}, \\
  \vec{p}'_2 &= \vec{p}_2 -\Delta \vec{p}.  \label{imper}
\end{align}
No torque acts at the points of contact, so that
\begin{eqnarray}
  \label{drstimp}
 \vec{\omega}_1^{'}&=\vec{\omega}_1
  +\frac{s_{12}}{I}\vec{u}_1\times \Delta\vec{p},
  \nonumber \\
  \vec{\omega}_2^{'}&=
  \vec{\omega}_2-\frac{s_{21}}{I}\vec{u}_2\times  \Delta\vec{p}.
\end{eqnarray}
holds.  To determine the collision process we consider the relative
velocity of the contact points which is given by
\begin{align}  \label{vrel}
  \vec{V}_r
  &= \frac{\vec{p}_{1}-\vec{p}_2}{m}+s_{12}\dot{\vec{u}}_1
-s_{21}\dot{\vec{u}}_2.
\end{align}
We introduce two parameters $\epsilon $ and $\beta$ to characterize
normal and tangential restitution:
\begin{alignat}{2}
  \vec{V}_r'\cdot \vec{u}_\perp &= - \epsilon \vec{V}_r\cdot
\vec{u}_\perp
  \label{epili} &\quad \epsilon &\in [0,1]  \\
  \vec{V}_r'\cdot \vec{u}_1 &= - \beta \vec{V}_r\cdot \vec{u}_1
  \label{betili1} && \\
  \vec{V}_r'\cdot \vec{u}_2 &= - \beta \vec{V}_r\cdot \vec{u}_2
\label{betili2}
\quad &\beta &\in [-1,1]
\end{alignat} 
The above equations characterize the collision process completely and
determine the postcollisional momenta in terms of the precollisional
ones.

In this paper we specialize to the case of perfectly smooth needles,
i.e. $\beta =-1$, which considerably simplifies the collision rules.
The change of linear momentum is then given by $\Delta \vec{p} =
\alpha \vec{u}_\perp $ with
\begin{equation}\label{alpha}
  \alpha = \frac{-(1+\epsilon) (\vec{V}_r\cdot\vec{u}_\perp)}
{2/m+(s_{12}^2+s_{21}^2)/I}
\end{equation}

\section{Analytical Theory}
The derivation of a pseudo--Liouville--operator for a system of N
colliding, hard rods proceeds similar to the case of hard spheres
\cite{noije97.2,ernst69,huthmann97}.  The rods are confined to a
3-dimensional volume $V$ and interact via a hard core potential. The
velocity of orientation $\dot{\mbox{\boldmath${u}$}}_1=\vec{\omega}_1
\times \vec{u}_1$ is confined to the plane perpendicular to
$\vec{u}_1$ and is therefore described by {\em two} generalized
canonical momenta, $p_{\theta_1}= I \dot{\theta_1}$ and $p_{\phi_1}
=I\dot{\phi_1}\sin^2\theta_1$, using spherical coordinates for the
orientation.  The total kinetic energy then reads
\begin{equation}\label{ham}
  {\cal{H}}_{\rm kin}
    = \sum_{i=1}^{N}\left(
  \frac{1}{2 M}\vec{p}_i^2+\frac{1}{2I}p_{\theta_i}^2+
  \frac{1}{2I\sin^2\theta_i} p_{\phi_i}^2\right).
\end{equation}
The time development of a dynamical variable
$A=A(\{\mbox{\boldmath$r$}_i(t),\mbox{\boldmath$u$}_i(t)
,\mbox{\boldmath$p$}_i(t),\dot{\mbox{\boldmath$u$}}_i(t)\})$ for
positive times is determined by the pseudo--Liouville--operator ${\cal
  L}_{+}$
\begin{equation}
   A(\{\mbox{\boldmath$r$}_i, \mbox{\boldmath$u$}_i  ,
\mbox{\boldmath$p$}_i,\dot{\mbox{\boldmath$u$}}_i\},t) =\exp(i{\cal
L}_{+}t)A(\{\mbox{\boldmath$r$}_i,\mbox{\boldmath$u$}_i,\mbox{\boldmath$p$}_i,
\dot{\mbox{\boldmath$u$}}_i\},0).
\end{equation}
The pseudo--Liouville--operator consists of two parts, ${\cal
  L}_{+}={\cal L}_0 +{\cal L}^{'}_{+}$.  The first describes free
streaming and can be expressed by the Poisson bracket with the kinetic
part of the Hamiltonian $ i {\cal L}_0 = \{{\cal H}_{\rm kin}, \ldots
\}_{\rm P.B.} $. The second, ${\cal L}^{'}_{+}$, describes binary
collisions of two hard rods
\begin{multline}
  \label{stossit}
 i {\cal L}'_{+} = 
   \frac{1}{2}\sum_{i\ne j}
     \left|\frac{d}{dt}|{\vec{r}_{ij}^\perp}|\right|
    \Theta(- \frac{d}{dt}|\vec{r}_{ij}^\perp|) \times \\
  \Theta(L/2-|s_{ij}|)\Theta(L/2-|s_{ji}|)
  \delta(|\vec{r}_{ij}^\perp|-0^+)
  (b_{ij}-1).
\end{multline}
Here $\Theta(x)$ is the Heaviside step function.  The interpretation
of the pseudo--Liouville--operator of eq. (\ref{stossit}) is
intuitively clear. The factor
$\left|\frac{d}{dt}|{\vec{r}_{ij}^\perp}|\right|$ is the component of
the relative velocity of the contact points perpendicular to both
rods.  It yields the flux of incoming particles. $ \Theta(-
\frac{d}{dt}|\vec{r}_{ij}^\perp|) $ is nonzero only if the two
particles are approaching and
$\Theta(L/2-|s_{ij}|)\Theta(L/2-|s_{ji}|)
\delta(|\vec{r}_{ij}^\perp|-0^+) $ specifies the conditions for a
collision to take place. The operator $b_{ij}$ replaces momenta and
angular momenta of particles $i$ and $j$ before collision by the
corresponding ones after collision.

The time evolution of nonequilibrium expectation values of an
observable
$A(\{\vec{r}_i,\vec{u}_i,\vec{p}_i,\dot{\vec{u}}_i\},t)=A(\Gamma;t)$
is defined by:
\begin{equation}
\label{expectation}
  \left \langle A \right\rangle_t=\int d\Gamma \rho (\Gamma;0) A(\Gamma;t)=\int d\Gamma \rho
  (\Gamma;t) A(\Gamma;0) \,.  
\end{equation}
$\Gamma$ denotes the whole phasespace and $\rho(\Gamma;t)$ is the
$N$-particle phase space distribution function, whose time evolution
$\rho(\Gamma;t)=\exp {(-i{\cal L}_+^{\dagger}t)}\,\rho(\Gamma;0)$ is
governed by the adjoint ${\cal L}_+^{\dagger}$ of the time evolution
operator ${\cal L}_+$.  Here we are interested in the average
translational and rotational kinetic energy per particle $E_{\rm tr} =
{m}/({2N}) \sum_i \mbox{\boldmath$v$}_i^2$ and $E_{\rm rot} =
{I}/({2N}) \sum_i \mbox{\boldmath$\omega$}_i^2$, as well as the total
kinetic energy $E=E_{\rm tr}+E_{\rm rot}$.

It is impossible to calculate expectation values as given in eq.
(\ref{expectation}) exactly and we are forced to approximate the
N-particle distribution function.  We assume that the system stays
spatially homogeneous and that both linear and angular momenta are
normally distributed.  In a system which is prepared in a thermal
equilibrium state the initial decay rates can be computed exactly and
yield different values for averaged translational and rotational
energy.  This suggests defining two different temperatures for the
translational and rotational degrees of freedom, corresponding to the
following ansatz for the N particle distribution function
\cite{huthmann97}
  \begin{equation} \label{rhohcs2}
 \rho_{\rm HCS}(\Gamma;t) \sim \exp \left [-
\frac{E_{\rm tr}}{T_{\rm tr}(t)} -
  \frac{E_{\rm rot}}{T_{\rm rot}(t)} \right ].
\end{equation}
$\rho_{\rm HCS}(\Gamma;t)$ depends on time via the average
translational $T_{\rm tr}(t)={2/3}\langle E_{\rm tr}\rangle$ and
rotational energy $T_{\rm rot}(t)=\langle E_{\rm rot}\rangle $. We are
interested in the cooling properties of a gas of hard needles and
compute the expectation values $\dot{T}_{\rm tr}={2/3}\langle i{\cal
  L}_+ E_{\rm tr}\rangle $ and $\dot{T}_{\rm rot}=\langle i{\cal L}_+
E_{\rm rot}\rangle $. Using the approximate many particle distribution
of eq. (\ref{rhohcs2}), we find two coupled differential equations
\begin{multline} \label{dgl1}
\frac{2\dot{T}_{\rm tr} }{\gamma T_{\rm tr}^{3/2}(1+\epsilon)} =
-\int_{\Box} d^2r \frac{(1+\frac{T_{\rm rot}}{T_{\rm
        tr}}kr^2)^{1/2}}{1+kr^2} \\
+\frac{1+\epsilon}{2}
\int_{\Box} d^2r\frac{(1+\frac{T_{\rm rot}}{T_{\rm
        tr}}kr^2)^{3/2}}{(1+kr^2)^2},
\end{multline}
\begin{multline}\label{dgl2}
\frac{4\dot{T}_{\rm rot} }{3\gamma T_{\rm tr}^{3/2}(1+\epsilon)} =
-\int_{\Box} d^2r \frac{\frac{T_{\rm rot}}{T_{\rm
      tr}}kr^2 (1+\frac{T_{\rm rot}}{T_{\rm
        tr}}kr^2)^{1/2}}{1+kr^2} \\
+ \frac{1+\epsilon}{2}
\int_{\Box} d^2r \frac{kr^2 (1+\frac{T_{\rm
        rot}}{T_{\rm tr}} kr^2)^{3/2}}{(1+kr^2)^2}
\end{multline}
with $\gamma = (2NL^2\sqrt{\pi})/(3V\sqrt{m})$ and $k=(mL^2)/(2I)$.
The two dimensional integration extends over a square of unit length,
centered at the origin.

\section{Simulations}
Simulations are performed using an event driven algorithm where the
particles follow an undisturbed translational and rotational motion
until a collision occurs. The velocities after the collision are
computed according to the collision rules eqs.
(\ref{imper})-(\ref{alpha}).  For this algorithm collision times for
each pair of needles have to be determined numerically, where we
follow the algorithm proposed by Frenkel and Maguire \cite{frenkel83}.
More efficiency is achieved by using the stratagem of Lubachevsky
\cite{lubachevsky91} and a linked cell structure, which allows us to
look for collision partners only in the neighborhood. The algorithm is
reasonably fast as long as there are only few needles in each cell of
the linked cell structure, so that the time consuming search for
collisions is restricted to the needles in the own and the neighboring
cells.  On the other hand we have to choose the linear dimension of
these cells to be larger than the length of a needle, so that for high
densities there are many needles in each cell and the algorithm
becomes slow.

We mention here that the algorithm runs into numerical problems if the
time between two collisions becomes too short to be resolved properly
as it usually happens during an inelastic collapse.  To circumvent
this problem, we use the $t_{c}$-model \cite{luding98}: if the time
between a collision and the preceding one for at least one particle is
smaller than a critical value $t_c$, $\epsilon$ is set to 1. We
believe that the influence of this procedure is small: there occurred
only two instances in the simulations presented here.

We performed simulations for various system sizes ($N={\mathcal
  O}(10^4)$) and coefficients of restitution in the regime of small
and moderate densities ($\frac{N}{V}L^3 \lesssim 1$). For high
densities ($\frac{N}{V}L^3 \gtrsim 10$) only a few simulations could
be done. We show here a simulation of $N=20000$ needles in a box of
length $12L$ with $\epsilon=0.9$.

\section{Hydrodynamic Quantities} 

To investigate deviations from {\it homogeneity}, we divide the
simulation box into cells whose linear dimension is chosen to be large
compared to the length of the needles but small compared to the box
size. Given the limitations due to finite size we choose cells such
that on average about 25 needles are in one cell. We then compute for
each cell the number density $\rho_\alpha= \frac{1}{V_{\text{cell}}}
\sum_{i \in \text{cell}_\alpha} 1=:\langle 1 \rangle_{\text{$\alpha$}}
$, the translational energy per particle $\rho_\alpha
E^{\text{tr}}_\alpha=
\langle\frac{m}{2}\vec{v}_i^2\rangle_{\text{$\alpha$}} $ and the
hydrodynamic temperature $T^{\text{tr}}_\alpha = E^{\text{tr}}_\alpha
- m \vec{U}_\alpha^2/2$.

{\it Momentum organization} of the particles shows up in the
hydrodynamic flow field $ \rho_\alpha \vec{U}_\alpha=\langle \vec{v}_i
\rangle_{\text{$\alpha$}} $. A good indicator\cite{brito98} for the
build up of long range velocity correlations is the ratio of the total
kinetic energy of the flow to the total internal energy $ S :=
(\sum_\alpha \frac{m}{2} \rho_\alpha \vec{U}_{\alpha}^2) /(\sum_\alpha
\rho_\alpha T_{\alpha}^{\text{tr}}) $.

To investigate {\it orientational ordering} we compute in addition the
quadrupolar moment of the needles $ Q_i^{\vec{e}} := (3(\vec{u}_i
\cdot \vec{e} )^2-1)$. The unit vector $\vec{e}$ is chosen either
along the direction of the particle's velocity or fixed in space.

\section{Small and moderate densities}

For densities such that $\frac{N}{V}L^3 \lesssim 1$, the system
remains homogeneous, orientationally disorderd and without long range
velocity correlations up to the longest observed time scales, i.e.
when the energy has decayed to $10^{-11}$ of its initial value.  To
check for spatial clustering, histograms of fluctuations of the local
density, velocity and translational energy were compared to those of
an elastic system but no significant difference could be observed.
Fluctuations, e.g. of the local density do not increase with time but
remain stationary, so that we can compare our approximate theory with
the simulations. We show here a simuation which has been performed for
$10000$ needles in a box of length $24$ L. This corresponds to a
density of $\frac{N}{V} L^3\approx 0.72$ or an average center of mass
separation $l/L\approx 1.1$ .

In this range of densities, cooling of a gas of hard needles proceeds
in {\em two stages}. First, there is an exponentially fast decay
towards a state which is characterized by a constant ratio of
translational and rotational energy. Second, there is an algebraically
slow decay of both, the translational and rotational energy, such that
their ratio remains constant. Both of these regimes are correctly
predicted by our approximate theory, based on the assumption of
spatial homogeneity.

\begin{figure}[hbt]
\center  \epsfig{file=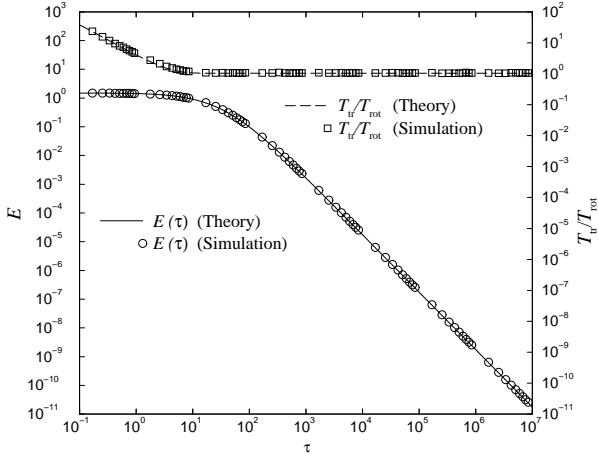, width=8cm}
  \caption{Double logarithmic plot of total kinetic energy $E = 3/2
    T_{\text{tr}} + T_{\text{rot}}$ in units of $T_{\text{tr}}(\tau=0)$ and of
    $T_{\text{tr}}/T_{\text{rot}}$ versus
    dimensionless time $\tau$. The simulation data are from a system 
 of 10000 needles, box of length 24 $L$ and $\epsilon=0.8$. }
  \label{bild2}
\end{figure}

In Fig. \ref{bild2} we plot the numerical solution of eqs.
(\ref{dgl1},\ref{dgl2}) for $\epsilon = 0.8$ and $k=6$ as a function
of dimensionless time $\tau = \gamma \sqrt{T_{\rm tr}(0)} t$. The
total kinetic energy $E = \frac{3}{2} T_{\text{tr}} + T_{\text{rot}}$
(in units of $T_{\text{tr}}(\tau=0)$) and the ratio
$T_{\text{tr}}/T_{\text{rot}}$ are compared to simulations.
Analytical theory and simulation are found to agree within a few
percent over eight orders of magnitude in time.  ($T_{\rm rot}(0)=0$
has been chosen as initial condition).
The decay of $T_{\text{tr}}/T_{\text{rot}}$ to a constant value
happens on a timescale of order one. In this range of times the total
kinetic energy $E$ remains approximately constant (on a logarithmic
scale) and decays like $t^{-2}$ only {\em after} translational and
rotational energy have reached a constant ratio.

\begin{figure}[hbt]
  \center  \epsfig{file=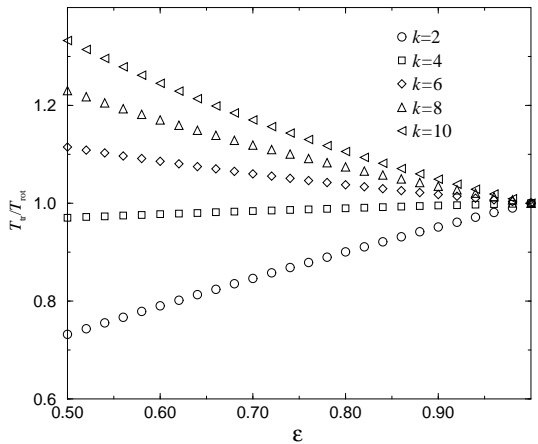, width=7cm, height=6cm}
  \caption{Asymptotic ratio $T_{\text{tr}}/T_{\text{rot}}$ as a
    function of $\epsilon$ and $k$.}
  \label{bild3}
\end{figure}

Eqs. (\ref{dgl1},\ref{dgl2}) allow for a solution with a constant
ratio of $T_{\rm tr}/T_{\rm rot}=c$ and both $T_{\rm tr}$ and $T_{\rm
  rot}$ decaying like $t^{-2}$. To determine the constant $c$, we plug
the ansatz $cT_{\rm rot}=T_{\rm tr}$ into eqs. (\ref{dgl1},\ref{dgl2})
and use $c\dot {T}_{\rm rot}-\dot {T}_{\rm tr}=0$.  This yields an
implicit equation for $c$, whose solution is plotted in Fig.
\ref{bild3} as a function of $k$ and $\epsilon$.
Setting $c=1$ in this implicit equation yields an equation for $k$
which reads
\begin{equation}
\label{kstern}
 (1-\epsilon^2) \int d^2x \frac{1-\frac{3}{2} k^*r^2}
  {\sqrt{1+k^*r^2}}=0,
\end{equation}
i.e. equipartition holds for {\em all} values of $\epsilon<1$ if
$k=(mL^2)/(2I)$ is set to the particular value $k^*=4.3607$, given as
the solution of eq. (\ref{kstern}). For $\epsilon=1$, equipartition
always holds, independent of $k$.

For $k<k^*$ we find $ T_{\rm tr} < T_{\rm rot}$ and for $k>k^*$ $
T_{\rm tr} > T_{\rm rot}$. Hence the distribution of mass along the
rods determines the asymptotic ratio of rotational and translational
energy, including equipartition as a special case.

The asymptotic solution discussed above is approached for arbitrary
initial conditions for long times.  If a totally elastic system is
prepared in an initial condition with $T_{\rm tr}\neq T_{\rm rot}$, we
expect that the equilibrium state (equipartition) is reached
exponentially fast with a relaxation rate given by $\nu\sim
\gamma\sqrt{E(0)}$.  As long as energy dissipation due to inelastic
collisions is small, we expect similar behavior, as indeed the
numerical simulations show.

\section{Dense systems}

\begin{figure}[htbp]
  \begin{center}
    \leavevmode
    \epsfig{file=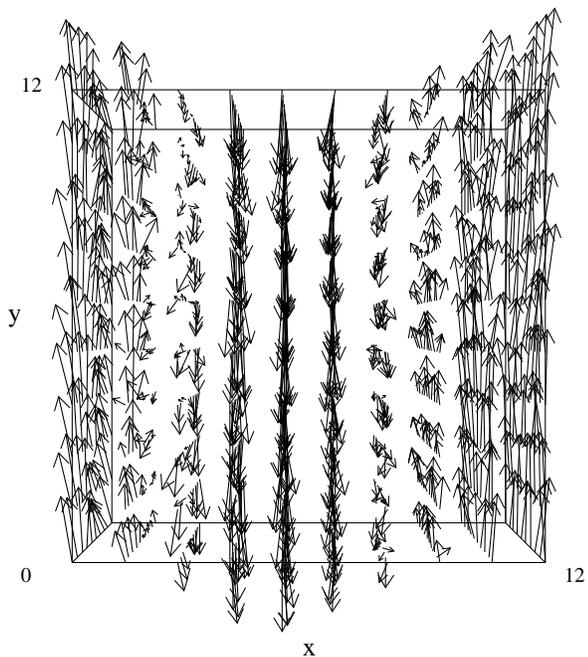,width=8cm}
  \end{center}
\caption{Flow field of the system (20000 needles in a box of volume
  $(12L)^3$ and $\epsilon = 0.9$) after 600 collisions per particle
  (the length of the velocity vectors are in arbitrary units).  }
\label{bild4}
\end{figure}

Simulations of inelastic hard spheres show well developed density
clusters and vortex patterns
\cite{goldhirsch93,mcnamara96,noije97,goldhirsch932}, if allowed to
evolve freely for sufficiently long times. The dominant mechanism for
the formation of vortex structures has been traced back to noise
reduction \cite{brito99}: After many collisions the particles move
more and more parallel. It is not clear a priori, whether such a
mechanism should also apply to rotating needles. In the simulations we
clearly observe the formation of large scale structures in the
velocity field.  In Fig. \ref{bild4} we show the hydrdynamic flow field
after 600 collisions per particle for a system of density
$\frac{N}{V}L^3\approx 11.6$, corresponding to an average center of
mass separation $l/L=0.44$.

\begin{figure}[htbp]
  \begin{center}
    \leavevmode
    \epsfig{file=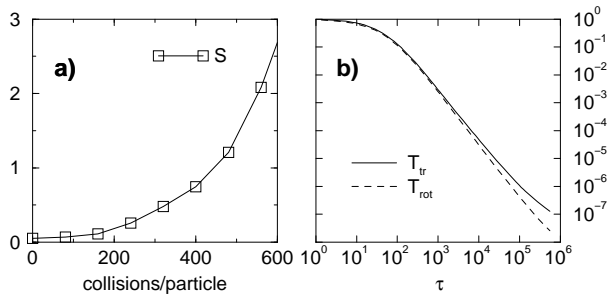,width=8cm}
  \end{center}
\caption{a) $S$ as a function of collisions per particle.
  b) Double logarithmic plot of kinetic energy per degree of freedom
  $T_{\text{tr}}$ and $T_{\text{rot}}$ in units of
  $T_{\text{tr}}(\tau=0)$ versus dimensionless time $\tau$.  }
\label{bild5}
\end{figure}

We observe two shear bands (note the periodic boundary conditions),
which move in opposite directions. Within a band the local flow field
is to a large degree aligned. The dominant part of the velocity of
each particle $\vec{v}_i $ is given by the flow $\vec{U}$ so that a
large fraction of the kinetic energy is in the flow and the ratio $S$
should be high. In Fig. \ref{bild5} a) we show $S$ as a function of the
number of collisions per particle. $S$ increases from a value of 0.05
to a value of about 2.5, i.e. by a factor of 50.

\begin{figure}[htbp]
\parbox{9cm}{\epsfig{file=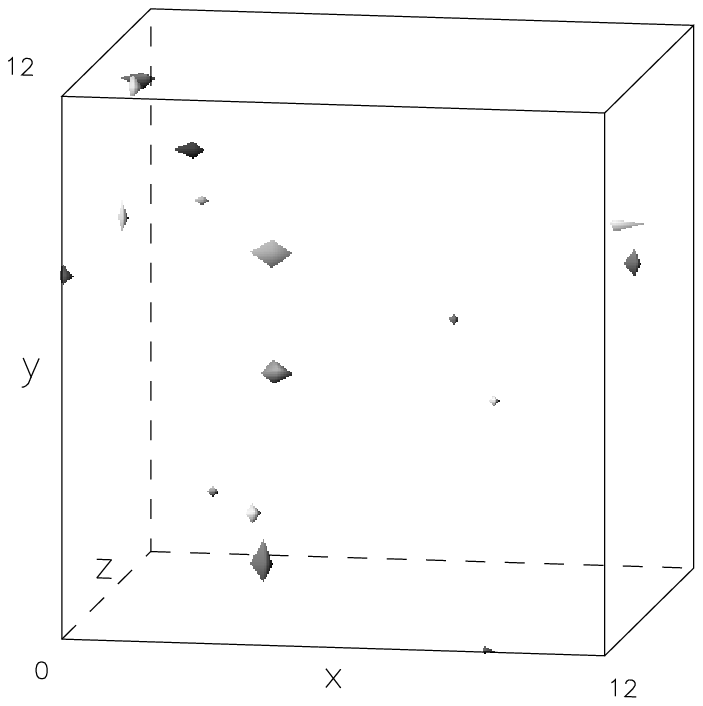,width=8.5cm}
\raisebox{1.5cm}{\makebox[0cm][r]{\bf a)\hspace*{8.0cm}} } }

\parbox{9cm}{\epsfig{file=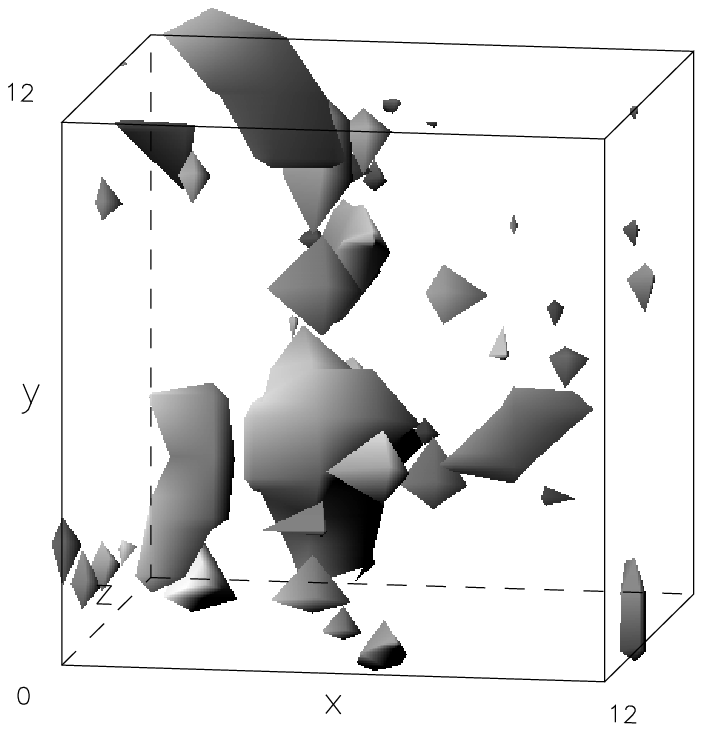,width=8.5cm}
\raisebox{1.5cm}{\makebox[0cm][r]{\bf b)\hspace*{8.0cm}} } }

\caption{a) Density at the beginning of the
  simulations.  Only very few and small regions have a more than 50 \%
  higher density than average.  b) Large regions with higher density
  have built up after 600 collisions per particle.}
\label{bild6}
\end{figure}

To visualize spatial inhomogeneities we plot in Fig. \ref{bild6}
regions with local deviation of the density $\Delta \rho_\alpha > 0.5$
at the beginning of the simulation and after 600 collisions per
particle. Obviously clustering occurs. To quantify this obseravtion we
have computed the second moment of the density fluctuations. It is
found to increase by a factor of 6 over its initial value after $400$
collisions. After about 500 collisions per particle it decreases
again, indicating that clusters form and dissolve again.

\begin{figure}[htbp]
  \begin{center}
    \leavevmode
   \epsfig{file=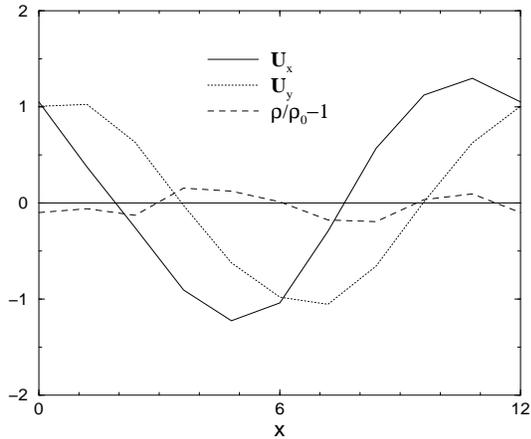,width=7cm,height=6cm}
      \end{center}
\caption{$y-$ and $z-$component of the flow field and fluctuation of the
  density as a function of $x$, for fixed $y=6L$, averaged over 10
  $z$-values.}
\label{bild7}
\end{figure}

For spheres one observes \cite{goldhirsch93,mcnamara96,goldhirsch932}
that most of the mass is concentrated in the two counterflowing
streams. To check for correlations between flow field and mass
density, we plot in Fig.  \ref{bild7} the components of the flow field
$\vec{U}_y$, $\vec{U}_z$ and the density fluctuation as a function of $x$, for
fixed $y=6L$ and averaged over 10 values of $z=1.2L,...,12L$. This and
similar plots give no hint of a strong correlation between flow field
and density fluctuations.

We also observe a weak tendency towards organisation of rotational
velocities.  The ratio of the kinetic energy of the local rotational
flow to the local rotational temperatur is found to increase by about
50 \% (as compared to an increase by a factor of $50$ for the
translational velocity).  Consequently, the deviation from Haff's
$t^{-2} $ cooling law is much stronger for the translational degrees
of freedom $T_{\rm tr}$ than for the rotational degrees of freedom
$T_{\rm rot}$ (see Fig. \ref{bild5} b)).

\begin{figure}[htbp]
  \begin{center}
    \leavevmode
  \epsfig{file=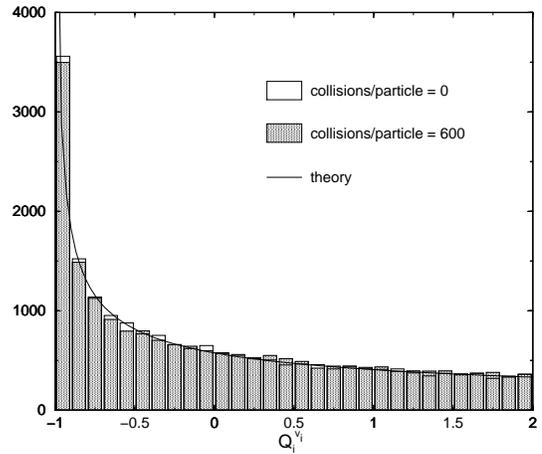,width=7cm } 
         \end{center}
\caption{Histogram of  $Q^{\vec{v}_i}$. 
  The distribution after $600$ collisions per particle coincides with
  the initial distribution and with the prediction for randomly
  distributed orientations.}
\label{bild8}
\end{figure}

To investigate alignment of the needles with the large scale velocity
flow field, we compute the quadrupolar momentum $Q^{\vec{e}}_i$ with
respect to the particle velocity, i.e.
$\vec{e}=\vec{v}_i/|\vec{v}_i|$. A histogram over all needles is shown
in Fig.  \ref{bild8}. The configuration after 600 collisions per
particle is compared to the initial state which corresponds to
randomized orientations. In addition we plot the theoretical
prediction for the histogram (straight line) which has been calculated
on the assumption that rods are oriented randomly and independent of
their velocity.
No indication for alignment of the needles can be  seen. Neither do we
observe a tendency for global ordering.

\section{Conclusion and outlook}

Our aim was to systematically investigate the cooling dynamics of a
granular system of nonspherical objects. We have chosen the simplest
nonspherical grains, hard needles, for which we were able to formulate
an approximate kinetic theory, generalising methods of kinetic theory
of elastic systems \cite{frenkel83} to inelastic collisions with
normal and tangential restitution. In addition, simulations of large
systems for various densities have been performed.

Analytical theory is so far restricted to the regime of moderate
densities, where the interparticle spacing is comparable to or larger
than the length of the needles. Such systems remain homogeneous and
show no long range correlations in the velocity field or orientation.
Cooling proceeds in two stages: 1) An exponentially fast initial decay
towards a state with constant ratio of translational to rotational
energy and 2) an algebraically slow decay, such that the above ratio
remains constant in time. The ratio of translational to rotational
energy is controlled by the coefficient of normal restitution and by
the distribution of mass along the rods.

Simulations in the dense regime, where the interparticle spacing is
smaller than the length of the needles, reveal large scale structures
in the translational velocity field. The density does not remain
homogeneous, but clusters form and dissolve again. Cooling proceeds in
three stages. For short and intermediate time scales the relaxation is
similar to the low density system, whereas on the longest time scales
we observe a crossover from the algebraic decay to an even slower
decay.  This latter decay may be identical to the asymptotic cooling
law of hard inelastic spheres, $\tau^{-d/2}$ in $d-$dimensions, where
$\tau$ is the average number of collisions suffered by a particle
within time $t$, which has been derived in \cite{brito98}.

We plan to generalise the hydrodynamic analysis to grains with
rotational degrees of freedom and in particular hard rods.  Another
possible extension of our work are rods of finite width and with
spherical endcaps.

{\em Acknowledgement}.--- This work has been supported by the DFG
through SFB 345 and Grant Zi209/5-1.

\end{document}